\documentclass[a4paper,12pt]{article}
\usepackage{epsfig}

\def\lsim{\mathrel{\rlap{
\lower4pt\hbox{\hskip-3pt$\sim$}}
    \raise1pt\hbox{$<$}}}     
\def\gsim{\mathrel{\rlap{
\lower4pt\hbox{\hskip-3pt$\sim$}}
    \raise1pt\hbox{$>$}}}     
\begin{document}
\title{Dynamical Interpretation  \\
   of Chemical Freeze-Out Parameters. }
\author{V.D.~Toneev$^{1,2}$, J. Cleymans$^3$,
E.G.~Nikonov$^{1,2} $, K.~Redlich$^{1,4}$,\\
and \\
A.A.~Shanenko $^{2}$\\
\vspace*{2 mm}
\small\em
$^1$Gesellschaft f\"{u}r  Schwerionenforshung, D-64220  Darmstadt, Germany \\
\small\em
$^2$ Bogoliubov Laboratory of Theoretical Physics,\\ 
\small\em
JINR, 141980 Dubna, Moscow Region,  Russia\\ 
\small\em
$^3$ Department of Physics, University of Cape Town, Rondebosch 7701, South Africa\\
\small\em
$^4$ Institute of Theoretical Physics, University of Wroclaw, PL-64291
Wroclaw, Poland.
}
\date{}
\maketitle

\begin{abstract}
It is shown that the condition for chemical freeze-out,
average energy per hadron
$\approx 1$ GeV, selects 
the softest point of the equation of state, namely the point where 
the pressure divided by the energy density, $p(\varepsilon)/\varepsilon$ has a minimum.
The  sensitivity to
the equation of state used is discussed. 
The previously proposed
mixed phase model, which is consistent with
lattice QCD data
naturally leads to the chemical freeze-out condition.
\end{abstract}


%
%

\newpage

Over the last few years the question 
of chemical equilibrium in heavy ion collisions has
attracted much attention \cite{sollfrank}. 
Assuming thermal and chemical equilibrium
within a statistical model, it has now been shown that
it is indeed possible to 
describe the hadronic abundances produced 
using  beam energies ranging 
 from 1 to 200 AGeV.
The
observation was made that the
 parameters of
the  chemical freeze-out curve  
obtained at CERN/SPS, BNL/AGS and GSI/SIS 
all lie on a 
unique  freeze-out curve in the $T-\mu_B$ plane. 
Recently, a surprisingly
simple interpretation of this curve has been proposed: the
hadronic composition at the final state is determined solely by an
energy per hadron of  approximately 1 GeV per hadron in the
rest frame of the system under consideration \cite{CR98,CR99}. 
In this letter
we propose a dynamical interpretation of the chemical
freeze-out curve and show that it is intimately related to the softest
point of equation of state defined by  the 
 minimum of the ratio $\displaystyle \frac{p}{\varepsilon}(\varepsilon)$ 
as a function of $\varepsilon$~\cite{HS94}. 
 Our considerations are essentially based on the
 recently proposed~\cite{NST98,TNS98}
mixed phase model which is consistent with the
available QCD lattice data \cite{karsch}.  
The underlying  assumption of the Mixed Phase (MP) model
is that  unbound quarks and gluons {\it may coexist}
with hadrons forming an {\it homogeneous}
quark/gluon--hadron phase \cite{NST98,TNS98}. Since the mean
distance between hadrons and quarks/gluons in this mixed phase
may be of same order as that between hadrons, their interaction
with unbound quarks/gluons plays an important role defining
 the order of the phase transition. 

Within the MP model \cite{NST98,TNS98} the
effective Hamiltonian is written in the quasi particle
approximation with the density-dependent mean--field interaction.  
 Under quite general requirements of  confinement
for color charges,
the mean--field potential of  quarks and gluons 
is approximated by the following form:
\begin{equation}
U_q(\rho)=U_g(\rho)={A\over\rho^{\gamma}}
\label{eq6}\end{equation}
with  {\it the total density of quarks and gluons} 
$$
\rho=\rho_q + \rho_g +\sum\limits_{j}\;n_j\rho_{j} 
$$
where $\rho_q$ and  $\rho_g$ are
the densities of unbound quarks and gluons outside of hadrons,
while $\rho_{j}$ is the density 
and $n_j$ is the 
number of valence quarks inside the hadron of type $j$. 
The presence 
of the total density $\rho$ in (\ref{eq6})
corresponds to the inclusion of the interaction
between all components of the mixed phase. 
The  approximation (\ref{eq6})  recovers two important
limiting cases of the QCD interaction, namely, if
$\rho \rightarrow 0$ the interaction potential goes to infinity,
i.e.  an infinite energy should be spent to create an isolated
quark or gluon  which ensures the confinement
of color objects and,  in the other extreme case of high energy density
corresponding to $\rho \rightarrow \infty$ we obtain the
asymptotic  freedom regime.

The use of a density-dependent  potential (\ref{eq6}) for quarks 
 and a hadronic potential described by a
modified non-linear mean--field model~\cite{Zim}  requires certain
constraints, related to thermodynamic consistency,
to be fulfilled~\cite{NST98,TNS98}. For the
chosen form of the Hamiltonian these conditions 
require that $U_g(\rho)$ and
$U_q(\rho)$ should be independent of the temperature. From
these conditions one also obtains an  expression for the
form of the quark--hadron potential~\cite{NST98}. 
  
A detailed study of the pure gluonic $SU(3)$ case with a  first
order phase transition allows one to fix the values of the 
parameters as $\gamma =0.62$ and $\displaystyle
A^{1/(3\gamma+1)} = 250\, MeV$. 
These values are then generalized to the 
 the $SU(3)$ system including quarks. 
For the case of quarks of
two light flavors at zero baryon density, $n_B=0$, the MP model 
is consistent with the
results from lattice QCD \cite{karsch}  with a deconfinement
 temperature $T_{dec}=153\;MeV$ and the crossover type of the
deconfinement  phase transition. The model can be  extended to
baryon-rich  systems in a parameter--free way \cite{NST98}. 

A particular consequence of the MP model is that for $n_B=0$
 the 'softest point' of the equation of state, as defined  
in~\cite{HS94}, is located at
comparatively  low values of
the energy density: $\varepsilon_{SP} \approx 0.45 \ GeV/fm^3 $.
This value of $\varepsilon $ is close to the energy density inside a
nucleon and, thus, reaching this value signals us that we are dealing
with  a single 'big' hadron consisting of deconfined matter. For
baryonic matter the softest point is gradually washed out at  $n_B \gsim 0.4 \
n_0$. As shown in~\cite{NST98,TNS98}, this behavior differs
 drastically  from both the interacting
hadron gas model which has  no soft point  and the 
two--phase approach, based on the bag model, having a first
order phase transition by construction and the softest point at
$\varepsilon_{SP} > 1 \ GeV/fm^3 $   independent
of $n_B$~\cite{HS94}. These differences should manifest
themselves in the expansion dynamics.

In Fig.1 we show 
 trajectories of the  evolution of central Au+Au collisions
in the $T-\mu_B$ plane  together with the freeze-out
parameters obtained from hadronic abundances. The initial state
was estimated using a transport model starting from a cylinder in the
center-of-mass frame with   radius
$R=4 \ fm$ and length $L=2R/\gamma_{c.m.}$ as described
in~\cite{NST98,TNS98}. The subsequent isoentropic
 expansion was calculated using a 
scaled hydrodynamical model with  the MP equation of state. 
As seen from the figure, the turning points of 
these trajectories correlate nicely
 with the extracted freeze-out parameters, as was noted
in~\cite{TNS98}, as well as with the smooth curve corresponding
to a fixed energy per hadron in the hadronic gas
model~\cite{CR98}.

The observed correlation is further elucidated in Fig.2. The quantity
$p/\varepsilon$ is closely related 
to the square of the velocity of sound  and
characterizes the expansion speed\footnote{In simple
hydrodynamic models, for example, the transverse expansion of a
cylindrical source, the evolution is governed by the
pressure-to-enthalpy ratio, $p/(p+\varepsilon)$~\cite{CGS86,rischke}.}, so
the system lives for the longest time around the softest point
which allows it to reach chemical equilibrium for the strongly
interacting components.
It is also seen that the position of the softest point 
correlates with the average energy per hadron being 
about  1 GeV in all nuclear
cases and even for  $p\bar p$ collision. One should note that
the quantity $\varepsilon /\rho_{had}$, where $\varepsilon$ is the
total energy density, coincides with
$\left<E_{had}\right>/\left<N_{had}\right>$
considered in~\cite{CR98} 
only in the case when there are no unbound quarks/gluons in the
system. In the MP model, all components are interacting with each
other and  therefore the quantity $\left<E_{had}\right>$ is not defined. The
admixture of unbound quarks at  the softest point
$\varepsilon_{SP}$  amounts to about  $13\%$ and $8\%$ at  beam 
energies $E_{lab}=150$ and $10 \ AGeV$, correspondingly.

  The MP equation of state plays a decisive role for the
regularity considered
here, describing both the order of the phase transition 
and the deconfinement temperature.  The
two-phase (bag) model exhibits a first order phase transition with
 $T_{dec}=160 \ MeV$ and has a {\it spatially
separated}  Gibbs mixed phase but the corresponding trajectories in the
$T-\mu_B$ plane are quite different from those in the MP
model as shown in~\cite{TNS98}. 
The exit point from the Gibbs mixed phase at
$E_{lab}=150$  AGeV is  close to the corresponding
freeze-out point in Fig.1. However the large differences 
noted above in 
$\varepsilon_{SP}$ and in its dependence on  $n_B$,  
mainly caused by the different type of the predicted 
phase transition, does not
lead to the observed correlation with the softest point position
in the whole energy range considered.
The interacting hadron gas model has no softest point effect as was
 demonstrated in~\cite{NST98,TNS98}. This fact is seen
also from Fig.2 where at $E_{lab}=2 \ AGeV$ the quark admixture is
practically degenerated ($\approx 1\%$) and instead of a  minimum
there occurs a monotonic fall-off specific for hadronic models with
a small  irregularity in $p/\varepsilon$ near  the point
$\varepsilon /\rho_{had}= 1 \ GeV$ \footnote{Note that at the
SIS energies the chemical freeze-out point practically coincides
with the thermal freeze-out~\cite{oeschler}}.

It is noteworthy that similarly to the results presented in Fig.2
the softest point of the equation of state correlates with 
an average energy per
quark, $\varepsilon/\rho \approx 350 \ MeV$ which is close to the
constituent quark mass. So, at higher values of 
$\varepsilon/\rho$ we are dealing with a strongly-interacting
mixture  of highly-excited hadrons and unbound massive
quarks/gluons forming (in accordance with Landau's idea~\cite{landau}) 
an 'amorphous'
fluid  suitable for hydrodynamic treatment.
Below the soft point the interaction deceases, the relative
fraction of unbound quarks/gluons decreases, higher hadronic
resonances decay into baryons and light mesons and thereby the value
of $\varepsilon/\rho$ goes down.  

In summary, the unified description of the chemical freeze-out
parameters found in~\cite{CR98} is naturally related to the fact that the
proposed condition
$\left<E_{had}\right>/\left<N_{had}\right> \approx 1 \ GeV$
selects the softest point of the equation of state where the strongly
interacting system stays for a long time. Such a clear correlation
is observed for the  equation of state 
of the mixed phase model but
not in purely hadronic  nor in two--phase models. 
In this respect the success of the MP
model in the dynamical interpretation of the freeze-out regularity
may be considered as an argument in favor of a crossover type
of the deconfinement phase transition in $SU(3)$ system with massive
quarks.

\vspace*{5mm}
We thank B.~Friman,  Yu.~Ivanov and W.~N\"{o}renberg for useful discussions.
E.G.N. and V.D.T. gratefully acknowledge the hospitality at the
Theory Group of GSI, where some part of this
work has been done. 
J.C. gratefully acknowledges the hospitality of the 
physics department of the University of Bielefeld.
This work was  supported in part by  
BMBF under the program of scientific-technological
collaboration (WTZ project RUS-656-96).

\newpage
{\bf {\large Figure captions} } \\

Fig.1.~The compiled chemical freeze-out parameters (borrowed
from~\cite{CR98,CR99})  obtained from the observed hadronic abundances
and dynamical trajectories calculated for
central $Au+Au$ collisions at different beam energies $E_{lab}$  with
the mixed phase equation of state. The smooth dashed curve is
calculated in the hadronic gas model for 
$\left<E_{had}\right>/\left<N_{had}\right> = 1
\ GeV$ ~\cite{CR98}. 

Fig.2.~The ratio of pressure to energy density, $p/\varepsilon$, versus
the average energy per hadron, $\varepsilon/\rho_{had}$, for
evolution of different systems.
The upper curve corresponds to $p\bar p$ collisions at
$\sqrt{S}=40 \ GeV$ with isoentropic expansion from a sphere
with $R=1 \ fm$. Other cases are calculated for central $Au+Au$
collisions at the given beam energy under the same conditions as in Fig.1. 
\newpage
%
%
%

\begin{thebibliography}{99}
%
\bibitem{sollfrank} For a review see e.g.  
J. Sollfrank,  J. of Physics G {\bf 23}, (1997) 
1903 and references therein.
%
\bibitem{CR98} J.~Cleymans and K.~Redlich, Phys. Rev.Lett. {\bf
81} (1998) 5284.
%
\bibitem{CR99} J. Cleymans and K.~Redlich, nucl-th/9903063.
%
\bibitem{HS94}	C.M.~Hung and  E.V.~Shuryak,
 Phys. Rev. Lett. {\bf 75} (1995) 4003; Phys. Rev. C {\bf 57} (1998)
      1891.
\bibitem{NST98} E.G.~Nikonov, A.A.~Shanenko and V.D.~Toneev,
Heavy Ion Physics {\bf 8} (1998) 89; nucl-th/9802018.
%
\bibitem{TNS98}  V.D.~Toneev, E.G.~Nikonov and A.A.~Shanenko,
 Proceedings of Interdisciplinary Workshop on Nuclear
Matter in Different Phases and Transitions, Les Houches, France,
March 30 - April 10, 1998, Kluwer Academic Publ. in Fundamental
Theories of Physics (to be published); Preprint GSI 98-30, Darmstadt, 1998.
%
\bibitem{karsch} 
See e.g. F. Karsch, Talk given at the ``Strong and Electroweak
Matter '98'', December 1998, Copenhagen, hep-lat/9903031.
%
\bibitem{Zim} Zimanyi J. {\it et al}.  Nucl. Phys.A{\bf 484} (1988) 647.
%
\bibitem{CGS86}
  J.~Cleymans, R.V.~Gavai and E.~Suhonen, Phys. Rep. {\bf 130} (1986) 217.
\bibitem{rischke} See e.g. D.H.~Rischke,
in  ``Hadrons in Dense Matter and Hadrosynthesis'', pp. 21-70,
Ed. J. Cleymans, H.B. Geyer and F.G.~Scholtz, Springer,  (1999).
%
\bibitem{oeschler} J. Cleymans, H. Oeschler and K. Redlich,
      Phys. Rev. C {\bf 59} (1999)  1663.
%
\bibitem{landau} L.D.~Landau, Izv. Akad. Nauk SSSR, {\bf 17}
(1953) 51.

\end{thebibliography}
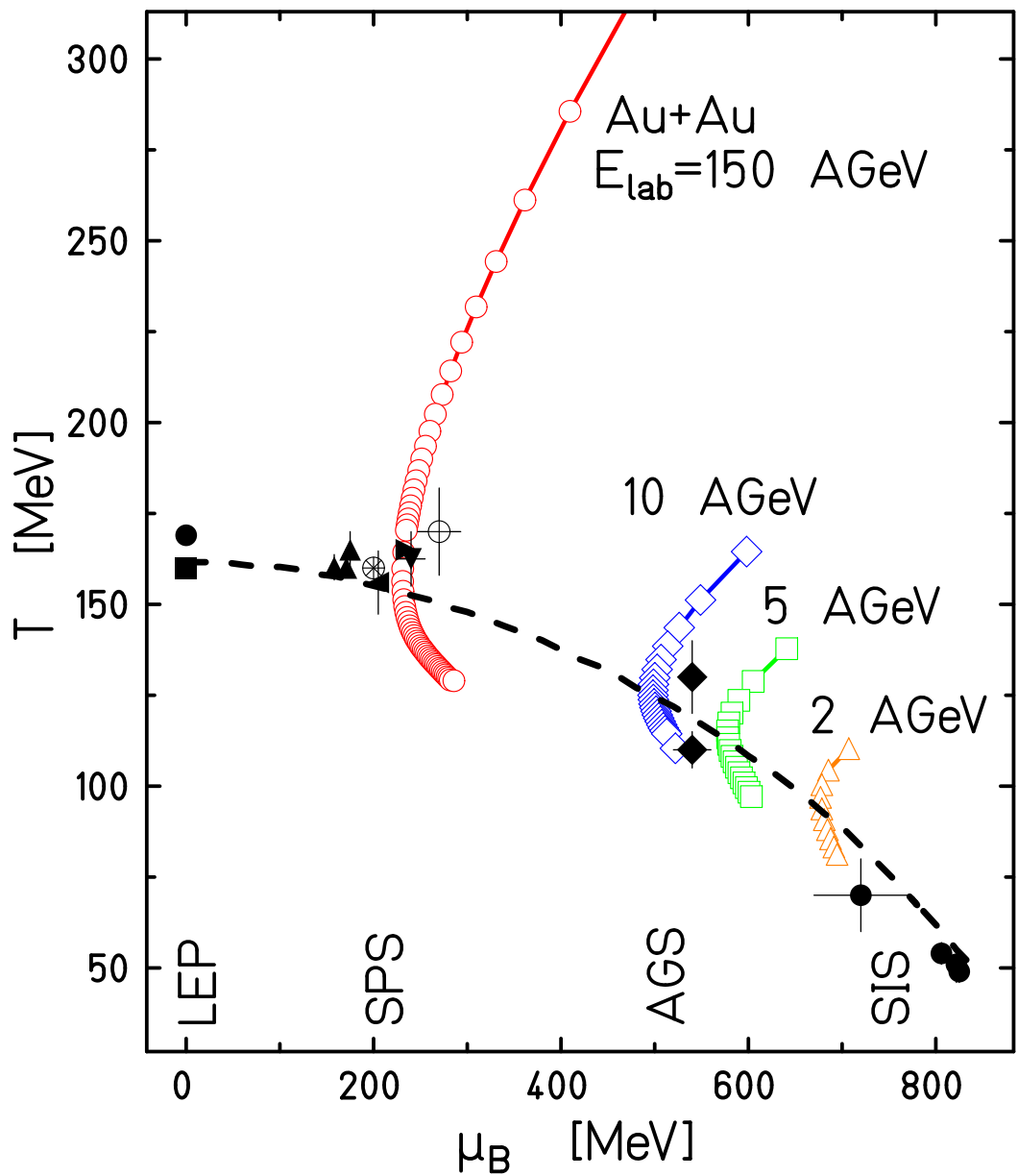
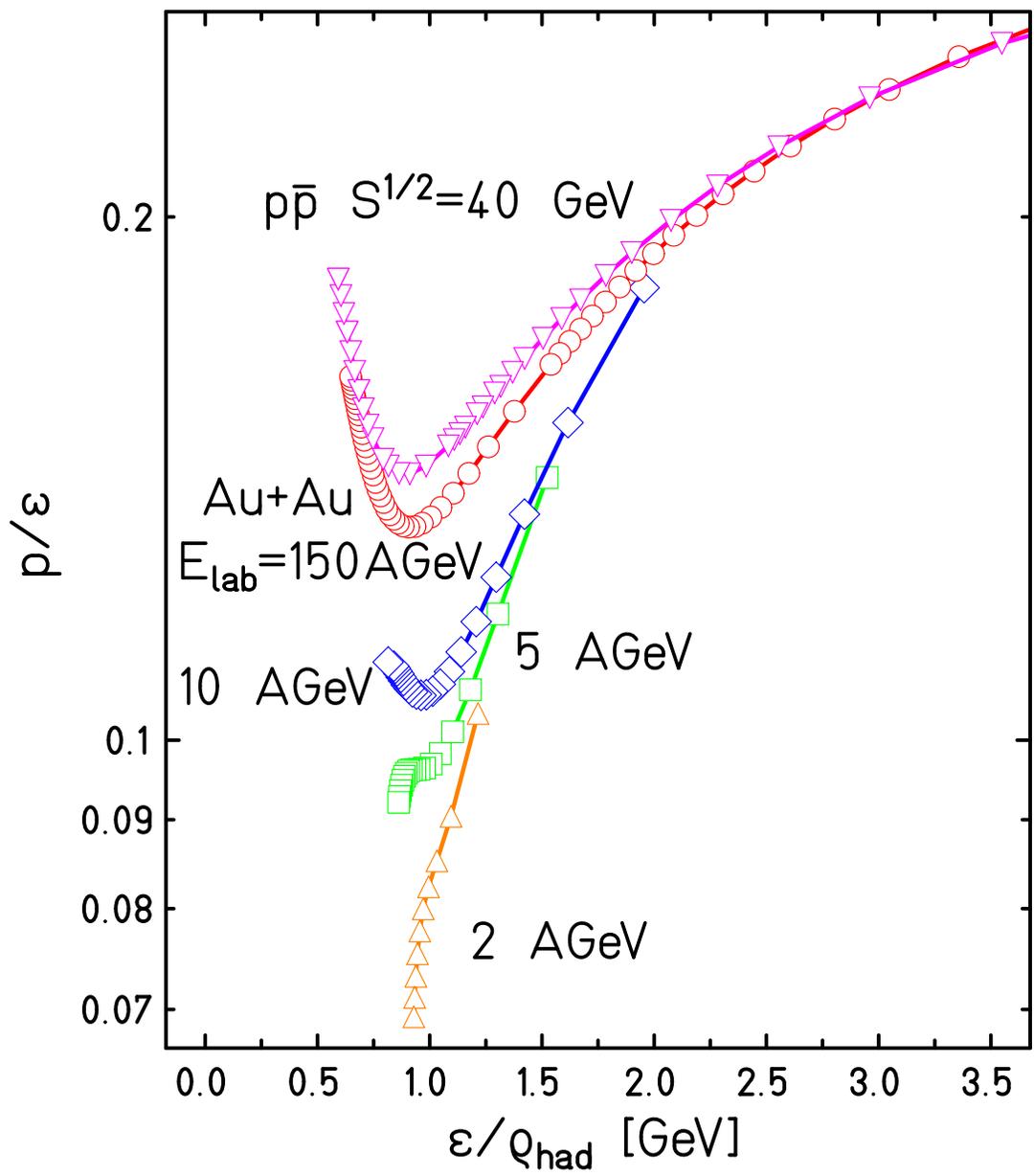
\end{document}